# KNOWLEDGE REUSE FOR CUSTOMIZATION: METAMODELS IN AN OPEN DESIGN COMMUNITY FOR 3D PRINTING[1]

**Harris Kyriakou**
IESE Business School, Av. Pearson 21, 08034 Barcelona, SPAIN {hkyriakou@iese.edu)

**Jeffrey V. Nickerson and Gaurav Sabnis**
Stevens Institute of Technology, Castle Point on Hudson, Hoboken, NJ 07030 U.S.A.
{jnickerson@stevens.edu} {gsabnis@stevens.edu}

*Theories of knowledge reuse posit two distinct processes: reuse for replication and reuse for innovation. We identify another distinct process,* ***reuse for customization****. Reuse for customization is a process in which designers manipulate the parameters of metamodels to produce models that fulfill their personal needs. We test hypotheses about reuse for customization in Thingiverse, a community of designers that shares files for three-dimensional printing. 3D metamodels are reused more often than the 3D models they generate. The reuse of metamodels is amplified when the metamodels are created by designers with greater community experience. Metamodels make the community's design knowledge available for reuse for customization—or further extension of the metamodels, a kind of reuse for innovation.*

**Keywords**: Knowledge reuse, metamodels, digital innovation, customization, parametric design, online communities, open source, software reuse, Thingiverse, 3D printing

## Introduction

Three-dimensional printing technology makes it possible to create physical objects by transforming digital files. This technology has the potential to revolutionize supply chains, because experts and novices alike can design, customize, and manufacture products locally for their own use (Balka et al. 2009; Kuk and Kirilova 2013; West and Kuk 2016). By contrast, traditional manufacturing processes are relatively inflexible and wasteful, because many identical objects are produced in far away places, transported at great expense to warehouses, distributed to retail outlets, and only then purchased by consumers (Gebler et al. 2014; Gershenfeld 2008; Raasch et al. 2009).

An important way that 3D printing technology is being diffused to consumers is through reuse of previously created designs. 3D printing communities are a lot like open source software communities: there is a culture of sharing and modifying designs through the editing of digital files (Fischer and Giaccardi 2006). What existing theories can be applied to these communities? Previous IS researchers have studied knowledge reuse (Allen and Parsons 2010; Markus 2001), and have distinguished between reuse for replication and reuse for innovation (Majchrzak et al. 2004). What is particularly salient in 3D printing, however, is customization, which has attributes of both replication and innovation. Specifically, in 3D printing communities it is possible to not only build models, but also to build more abstract models called metamodels; each abstract model can generate many concrete models. In practice, community members interactively modify each metamodel to produce many different 3D models. These 3D models are complete specifications that can be sent to 3D printers. Figure 1 shows the printed objects that resulted from the modifications of one metamodel.

[1]Sastish Nambisan, Kalle Lyytinen, Ann Majchrzak, and Michal Song were the accepting senior editors for this paper. Nicholas Berente served as the associate editor.

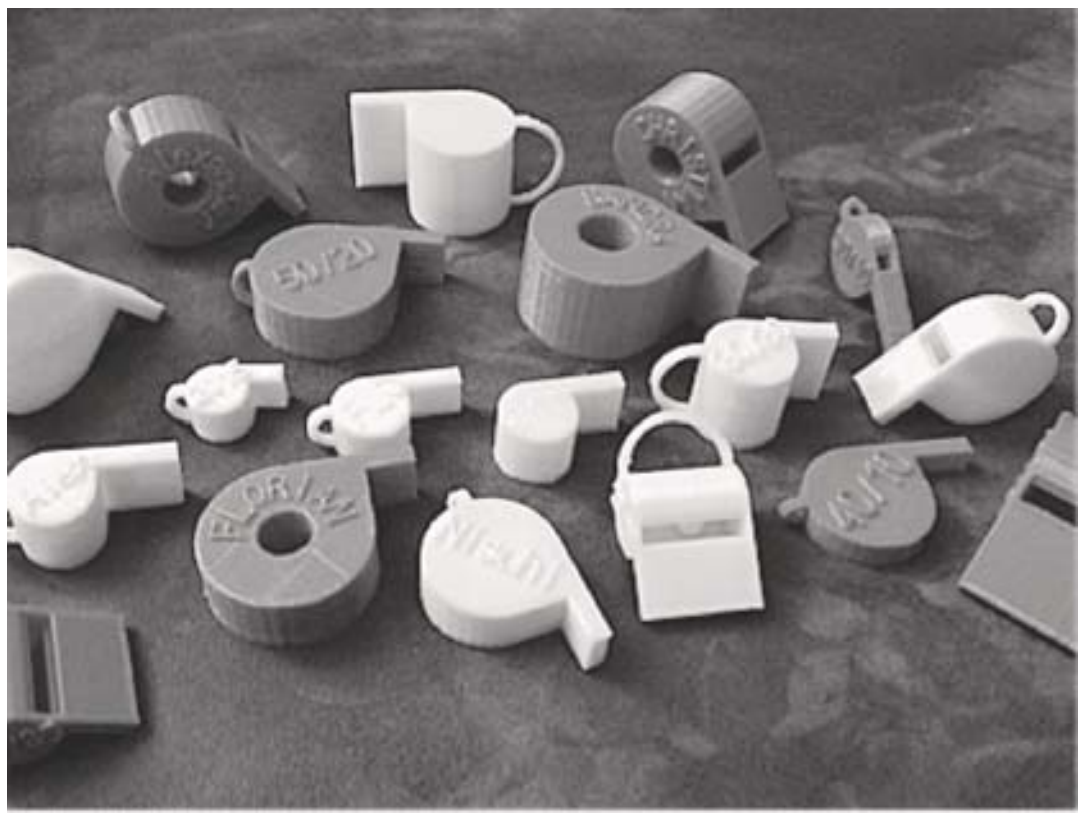

**Figure 1. Examples of 3D Printed Whistles Created from a Single Metamodel**

By themselves, customization, open source repositories, and 3D printing are not new phenomena. But the combination of these technologies and processes is novel. While it has been possible for years to customize a car or a computer, the choices have been limited. Consumers have rarely, if ever, been able to manipulate the continuous parameters of a design; instead, they choose from a small number of predefined options. In the context we discuss here, it is possible for community members to produce truly customized physical objects. As has been learned in studies of computer-assisted software engineering tools (Banker and Kauffman 1991), the introduction of meta-level tools can scaffold learning for novices. In a community, experts can create tools that novices can operate (Fischer and Giaccardi 2006). Thus, more people can participate in design activity, and as they do, they become more expert; eventually, they are no longer consuming other's designs, but generating their own. Such tools can also lower the cost of product variation because the design space, and the tools that manipulate the space, can be structured to streamline exploration (Yumer et al. 2015). We suggest that a systematic examination of metamodels in the 3D printing context can add to our understanding of how digital innovation can change design processes, manufacturing processes, and supply chains. This research note is a first step in that direction.

Our study is based on an examination of Thingiverse, currently the largest open design community dedicated to 3D printing. Through an analysis of the digital artifacts—the files and comments—shared in this community, we show the effects on reuse of a technology that allows metamodels to be modified to create customized designs. We find that metamodels are highly reused, but that the models subsequently generated by the metamodels are not reused as much. Metamodels created by experts are reused more than those created by novices.

The remainder of the paper is structured as follows. The next section provides a discussion of theories related to reuse and metamodels, culminating in our hypotheses. An example of these practices is described and is followed by tests of the hypotheses. We conclude with a discussion of the results and their implications for theory and practice.

## Theoretical Development

### *Reuse and Metamodels*

Knowledge reuse is the process by which previously created knowledge is repurposed, modified, and recombined (Alavi and Leidner 2001; Markus 2001). Majchrzak et al. (2004) identified two types of reuse. *Reuse for replication* is a form of knowledge acquisition: a particular problem needs solving, and knowledge is reused to solve that problem. No broader integration is needed, and no novelty results. By contrast, *reuse for innovation* involves integrating new knowledge with other knowledge, new and old, resulting in novelty. In open source environments, knowledge is shared with the community for further reuse (Howison and Crowston 2014), and both types of reuse occur.

Are there signs that guide reuse choices? One place to look is in the structures, processes, and data referred to as metaknowledge. Metaknowledge is knowledge about knowledge (Aiello et al. 1986), and has been shown to affect reuse (Evans and Foster 2011; Majchrzak et al. 2004).

In the domain of open design, some varieties of metaknowledge are formally instantiated in metamodels. A metamodel offers a language to define another language and thus what it can represent (Jarke et al. 2009); the roots of metamodels lie in the study of logic (Tarski 1983). They have broad application in engineering, architecture, manufacturing, and computing, because of the wide variety of the models they describe (see Frazer 2016; Kelly and Tolvanen 2008; Simpson et al. 2001).

In the context of engineering, a metamodel contains enough information to generate a range of related models, a family of designs, which, when printed, generate a family of products or parts. Indeed, reconfigurable, cellular, and additive manufacturing encourage a process of design utilizing metamodels that focus on designing not just one object, but a family of related objects (Koren and Shpitalni 2010; Tseng et al. 1996). These metamodels are sometimes called configurators, toolkits, and codesign platforms (Franke 2016; Jeppesen 2005; Piller and Salvador 2016). The Thingiverse community refers to these metamodels as *customizers.*

In the open design context being studied here, 3D metamodels allow designers to generate 3D models, which are descriptions of the surface geometry of an object (Woodbury 2010). These models are represented as data files that can be automatically converted into instructions for 3D printers, which in turn produce physical objects such as whistles. A 3D metamodel is a form of domain-specific modeling often practiced in software design (Kelly and Tolvanen 2008); a 3D metamodel is used to generate new 3D models just as a domain-specific programming language can be used to generate new software programs. We will from here on refer to 3D metamodels as *metamodels*, and 3D models as *models*.

The metamodels used in design practice consist of several integrated components. The first component is a set of parameters: these are the parameters that control the variety of the designs that will be generated. These parameters are bound to the second component, a model with named parameters. This is the template. The third component is an interface that allows a designer to manipulate the parameters by dragging slider bars or typing in textual information, a technique that emerged from human–computer interaction research metamodels. We are assuming the conjunction of all these components, as shown in Figure 2.

A user can assign values to a set of parameters; this leads to the production of a model, which the user can see and further adjust through modification of the parameter values (Figure 2). Through iteration, the user can systematically explore a design space. Figure 3 shows the results of modifying the radius and blower length of a whistle using a metamodel on Thingiverse.

From an information systems theory perspective, metamodels are made possible through three attributes of digital artifacts. Digital artifacts are *malleable*, which means they can be easily edited and recombined (Henfridsson and Bygstad 2013). They are also *interactive*, meaning that digital tools can allow users to see the effects of editing in real time (Henfridsson and Bygstad 2013). Finally, they are *reflexive*, in that they can refer to themselves (Kallinikos et al. 2013).

From a practical perspective, metamodels have proven effective in many consumer situations. For example, Adidas embeds knowledge about classes of sneakers in an application that allows consumers to design their own personalized sneakers, using a graphical user interface on a website (Piller et al. 2004). This is an example of consumer co-creation. Generally, metamodels provide a way to address heterogeneous preferences; they provide consumers an alternative to products created by dominant design strategies (see Abernathy and Utterback 1978).

In the context of open design, metamodels can be reused in two ways. First, parameters can be chosen, and the metamodel outputs a new model; this is what we refer to as *reuse for customization*. Second, the source code of the metamodel can itself be modified or recombined with another metamodel's source code, creating a new metamodel that can generate an extended or different family of models compared to its precursor. By contrast, models themselves represent a single design, not a design family, and for this reason may not have as broad an appeal. Moreover, studies of diffusion of innovation (Davis 1989; Rogers 2010) suggest that the ease of use incorporated into the metamodel interfaces will increase adoption. Metamodels are explicitly created with the intention of reuse for customization, while models are not (Tseng et al. 1996). This leads to the metamodel hypothesis (H1):

> H1: *Metamodels are more likely to be reused than models.*

## Generated Models

When metamodels are run, they produce models that are specific instances of the parameters of a metamodel. These *generated models* may be pleasing to their designers, but not to other members of the community who do not share the same specific needs. For example, a watchband customized with someone's initials is inherently uninteresting to almost everyone else. Even if a customized design is close to what someone else wants, it is more likely that one will use the metamodel instead of the generated model. Instead of taking a watchband customized with someone else's initials, deleting those initials, and putting their own initials in, they will find it easier to manipulate the metamodels and put in their own initials. By contrast, when confronted with models that don't have an associated metamodel, designers will reuse the model, because they don't have recourse to the metamodel.

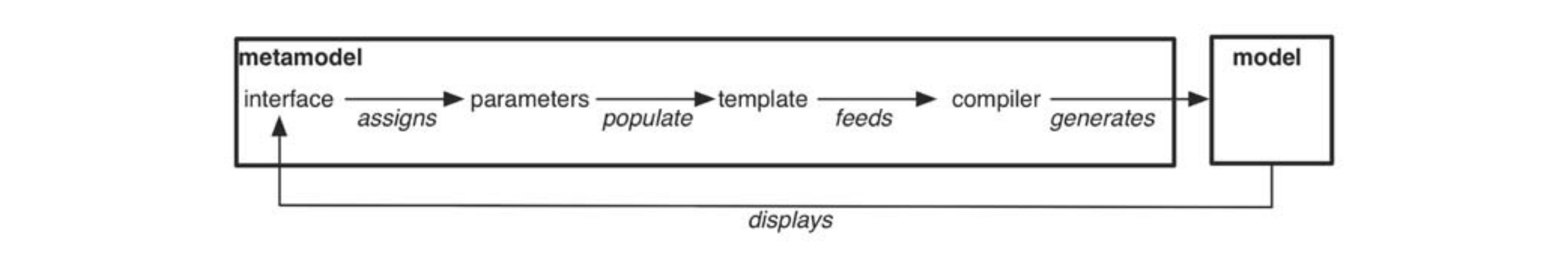

**Figure 2. Metamodel Components and Their Relationships**

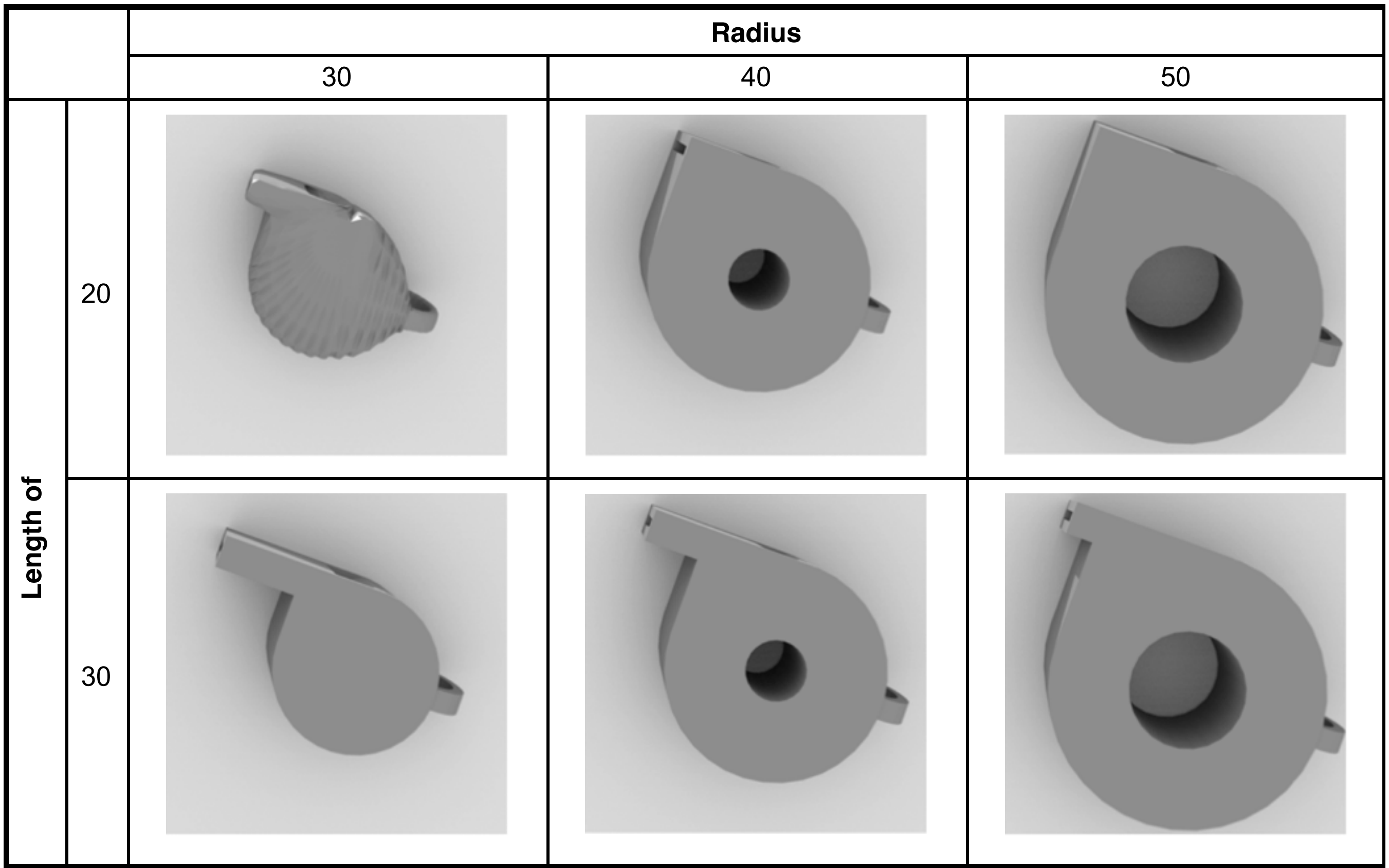

**Figure 3. Family of Designs Created by Varying the Radius and the Blower Length of the Whistle Using a Metamodel**

Theories of customization have found that both ease of use and hedonic motivations make metamodels attractive to users (Fogliatto et al. 2012). It may be easier to change the slider on a metamodel than to load a generated model into an editing tool. And it may be easier to find the metamodel than to search through the myriad generated models given that the likelihood of finding an exact match is low.

In sum, expertise-seeking novices are more likely to go back to a metamodel to generate a new model rather than work off an already generated model. And expert designers are more likely to work off the source code of the metamodel, extending it, rather than working off an informationally impoverished generated model. In other words, designs without metamodels will be reused because there is no other choice. Metamodels will be edited because such edits are productive, affecting entire design families. But generated designs are unlikely to be edited, because the metamodel is a more attractive starting point. This leads to the generated

model hypothesis (H2):

> H2: *Designs that are generated from metamodels are less likely to be reused than other designs.*

### The Designer and the Metamodel

Members of communities that share work are often experts in that they competently and reflectively practice their skills (Schön 1983). Communities also contain novices who are seeking or building expertise (Markus 2001). Experience has been shown to be an important driver of activity in open source communities: experts create reusable components, and novices reuse them (Lim 1994). Experience affects the use of certain programming language processes and technologies, as well as overall productivity (Haefliger et al. 2008; Mallapragada et al. 2012; Von Krogh and Von Hippel 2006).

The metamodel hypothesis (H1) suggests that the availability of a metamodel is positively related to the likelihood of design reuse, as the metamodel gives designers (novices and experts alike) a way to build on the knowledge of previous designers by encoding it in a malleable form. Although the metamodel enables reuse of existing knowledge, another driving force behind reuse is likely to be the quality of the knowledge itself, which comes from the designers. So the higher the quality of knowledge embedded in the metamodel, the more attractive it will be to users. This suggests a moderating effect: the metamodel makes it easier for the knowledge of expert designers to be reused by the community. Thus, we expect that previous community experience will positively moderate the positive relationship between metamodels and design reuse.

This should then lead to an interaction between the traits of the designer and the design format chosen, as stated in the designer experience hypothesis (H3):

> H3: *Metamodels will exhibit amplified reuse when created by members with higher levels of community experience.*

### Similarity in the Design Space

There is a long tradition of conceptualizing innovation as a search through the design space, a space in which changes in dimension produce new designs and points in large dimensional spaces (Brooks 2010; Frenken 2006; March 1991; Simon 1996). For example, designers of whistles can choose the overall shape of the whistle and the shape of its openings. While there are is an almost infinite number of choices available across these two dimensions, in practice, only certain combinations of choices will result in sound; a designer shapes the whistle, tests, and shapes again, exploring the space. An individual designer alternates between changing form and testing function (Frenken 2006).

Participants in open design communities can collectively explore the design space, potentially accelerating innovation. Because all designs are visible and all contribution history is available, it is possible to monitor the evolution of designs. It is also possible to understand the effect of similarity between designs, because we know at each point in time if a design was imitative or novel in relation to the preceding designs. This means designers may be able to make better choices about what innovations to pursue. Designs that are identical to parent designs are less likely to be reused because the knowledge embedded in them is already available. In contrast, designs that are dissimilar to their parent designs may further design space exploration, but may be considered marginal and may be ignored. In such a conceptualization, replication manifests as identical points, incremental innovation manifests as close points, and radical innovation manifests as far points in the design space.

Designs that were the result of *reuse for replication* processes will thus be more similar to the parent design than designs from *reuse for innovation* processes. As metamodels constrain reuse to a specific design space defined by their parameters, we expect them to yield designs that are similar to them. This leads to the design similarity hypothesis (H4):

> H4: *Metamodels are more likely than models to lead to designs similar to themselves, and therefore are less likely to lead to dissimilar designs.*

## Study Context: Open Design and Thingiverse

### 3D Printing and Thingiverse

Internet-based technologies have spurred the creation of digital communities, where knowledge is openly shared, but almost always remains in digital form. The RepRap project is an exception. Created in 2005 and still ongoing, it seeks to create 3D printers that replicate themselves by printing more 3D printers (Jones et al. 2011).

One of RepRap's core members, Zachary Smith, created Thingiverse in 2008 (Jones et al. 2011). It was intended to be an open design community where designers could freely download user-generated designs, or create something new by reusing existing designs and uploading their versions. Thingiverse, owned by MakerBot Industries, grew rapidly; it had more than 11,000 designs in June 2013, and more than 1,400,000 as of April 2016. By default, designers license their creations under a Creative Commons Attribution license. While designers can change the default, they tend not to: more than 98% of the designs are open in the sense that designers do not retain ownership rights.

MakerBot initially built their printers based on open designs. Later, they patented some of their technology, which upset their early supporters. A thorough and interesting discussion of the site's history can be found in West and Kuk (2016). While there are other repositories emerging, many also sponsored by 3D printing companies, Thingiverse remains by far the largest public design repository at this time, and for that reason was our choice for analysis.

Advances in technology have been increasing access to 3D printing. A 3D printer in the mid 1980s cost $100,000 or more (Hoffman 2016). Thirty years later, desktop 3D printers cost $200 or more, a 500-fold reduction in price (McMenamin et al. 2014). At a larger scale, the global market for 3D printers and services has been projected to grow from $2.5 billion in 2013 to $16.2 billion in 2018 (Earls and Baya 2014). Currently, 3D printers can produce objects in a variety of materials, including plastics, metals, and ceramics. Applications include the manufacturing of prosthetics (Rengier et al. 2010), buildings (Campbell et al. 2011), guns (Wohlers and Caffrey 2013), food (Tibbits 2014), human tissue (Mironov et al. 2003), and medicine (Schubert et al. 2013).

## File Types and Tools

Most contributions to Thingiverse are in one of two standard 3D model formats: STL for surface models and OpenSCAD for solid models. A third type of contribution, a metamodel, is created by inserting comments into OpenSCAD files that include parametric and user interface information. We describe these formats in more detail, because the formats and tools can have an effect on reuse.

STL is a stereolithography CAD file format. STL files describe only the surface geometry of 3D objects. They are edited using GUI-based CAD modeling tools such as Blender (Flavell 2010). OpenSCAD, however, is a text-based, programmer-oriented solid modeling tool that can be used to express models and convert these models into other formats, including STL. OpenSCAD files are similar to open source software files, because they are human readable and have attributes of scripting language syntax and semantics, including variables, conditionals, and subroutines. The language is free, released under the GNU General Public License, and is developed and distributed on GitHub (Kintel 2015). Figure 4 shows examples of different underlying file formats and the ways they are visualized.

STL files just record the vertices of triangle facets. While they theoretically could be edited by hand, it is nearly impossible to do so. OpenSCAD files look like a scripting language; it is possible to edit the text file in any text editor. Metamodels are OpenSCAD files with embedded parameters and interface bindings. The line "rad = 30; // [20:50]" is interpreted as follows: A parameter called *rad* for radius is set to 30 by default, with an allowed range between 20 and 50. It will have a slider bar associated with it, which can be seen in the screenshot of the interface.

Reuse in Thingiverse forms a network. Each contribution can have zero or more parent designs. These parents are edited using a tool, which produces a new design in a potentially different format. Figure 5 shows that the tools and designs form a bipartite network. Tools can transform types: in the third panel of Figure 5, an OpenSCAD design is transformed into a metamodel through a text editor.

### The Designers

As in many open innovation communities, Thingiverse members have a wide range of backgrounds and levels of expertise. The community includes professional designers, artists, programmers, educators, hobbyists, and curious novices. Designers use different editing software based on their background and the task at hand. Programming-savvy designers, aiming to create more industrial related designs such as cases, gears, and tools, are more likely to use OpenSCAD, while other designers edit designs using Blender in a process similar to shaping clay. Such software is preferred in tasks such as designing creatures, human figures, and faces. As an example, 5% of the designs in the fashion category included an OpenSCAD file compared to 17% in the 3D printing parts category. The community differs in a few ways from open source software communities. There is no concept of teams. Individual designers reuse each other's work. In open source software communities, the eventual product is a working program, whereas in Thingiverse, the eventual product is a 3D object in the physical world. Unlike software developers, designers need to worry about temperature, gravity, smoothness, adhesiveness, and many other physical properties. Knowledge shared within open design communities lies at the intersection of the digital and the physical, and reused knowledge often takes form in the physical world.

**STL Model**

```
facet normal 3.22623e-016 -1 2.65745e-016
    outer loop
      vertex -34.8211 -2 11.2533
      vertex -34.8211 -2 8.74667
      vertex -34.1978 -2 6.31875
    endloop
  endfacet
```

**SCAD Model**

```
module pfeife(name,sizename){
 if (rad > 39) {
    color("yellow")   writecylinder(name,[0,0,0],t=3,h=9,font="write/Letters.dxf",
      space=1.2, rad/2+1,hoehe,face="top");}
```

**Metamodel**

```
// Radius of the whistle in mm
rad = 30; // [20:50]
// Height of the whistle in mm
hoehe  =  20;  // [15:30]
// Textsize on whistle
textsize =  10;  // [8:14]
// preview[view:south, tilt:top]
```

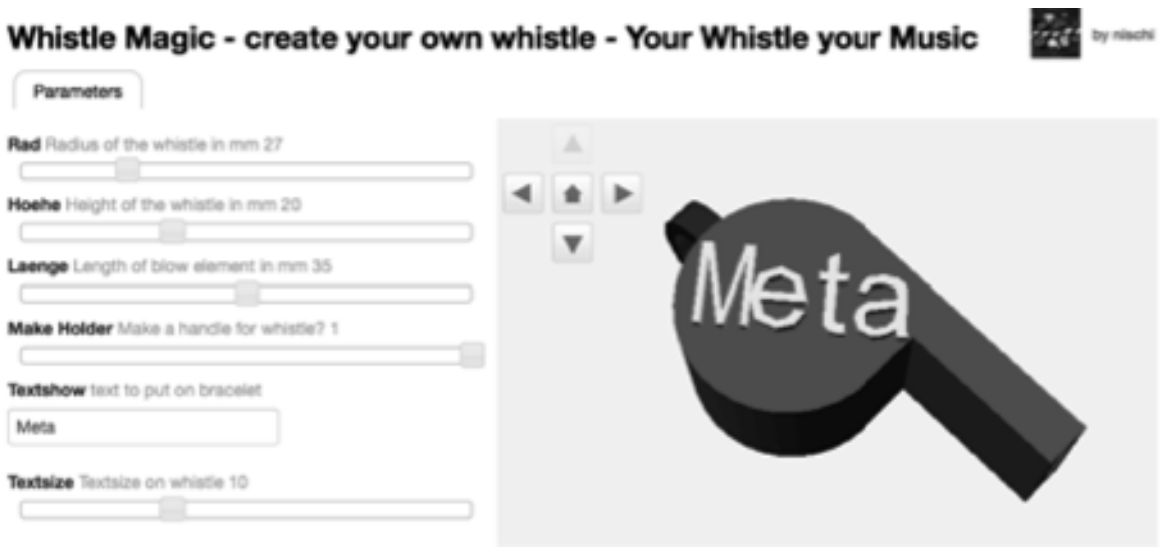

Metamodel Interface

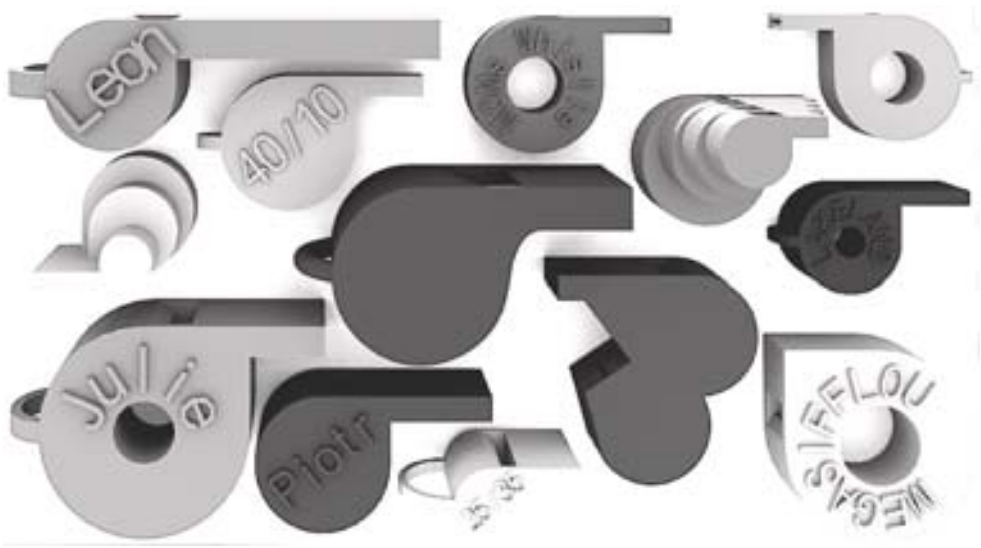

3D Printed Customized Designs

**Figure 4. File Types, Source Code, and the Metamodel Interface**

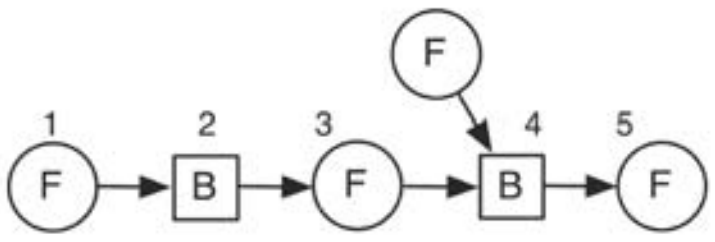

(1) An STL facet file (1) is edited using Blender (2) producing another STL file (3). Another designer recombines two models (4) into a third (5).

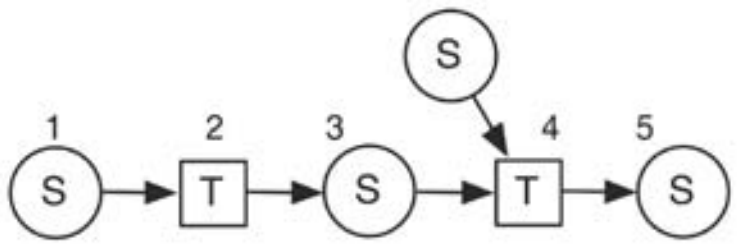

(2) An OpenSCAD solid model (1) is edited using text editor (2) producing another OpenSCAD Model (3). Another designer recombines two models (4) into a third (5).

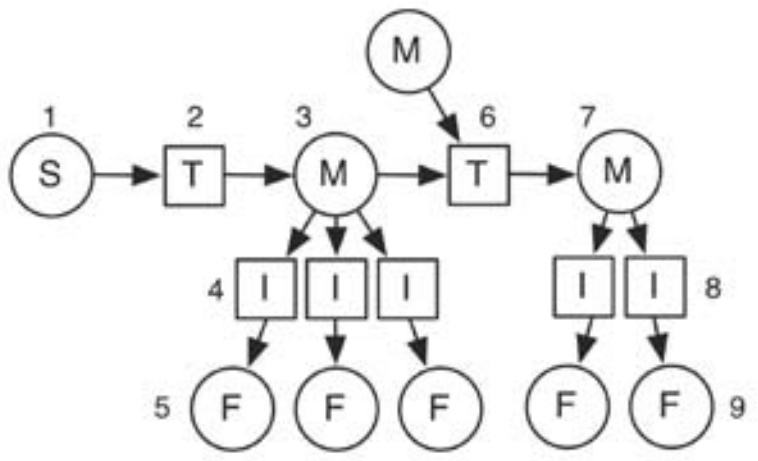

(3) An OpenSCAD design (1) is edited using text editor (2) into a metamodel (3). Designers run the metamodel interface (4), producing customized models in STL (5). Another designer recombines two metamodels in a text editor (6) to form a new one (7), which is customized (8,9).

F = STL facet file, S = OpenSCAD, M = Metamodel, B = Blender, T=Text editor, I = Interface

**Figure 5. Different Reuse Trajectories for (1) STL, (2) OpenSCAD, and (3) Metamodel**

## Research Design and Methodology

### Sample

We extracted data from Thingiverse using its application program interface (API). Given our focus on reuse for customization, we collected all designs created between January 7, 2013, the date of the inception of metamodel tools in Thingiverse, and June 2, 2013. The approximately five months of data provided us a long enough window to see chains of reuse, but not so long a window that our analysis would need to control for temporal changes or regime shifts in the platform or context. The number of designs collected was 24,173, which provided sufficient statistical power.

Our dependent variable, *Reuse,* was measured by counting the number of times each design was modified in the above time frame. Designers who reuse or recombine designs indicate it by linking their derived design to the parent design or designs, similar to how academic papers cite other papers. Self-reuse instances were not counted, for the same reasons self-citations are often excluded from measures of scholarly impact (Hyland 2003).

Hypothesis 4 included as dependent variables two measures of similarity between parent designs and the designs that reused them. Dissimilarity, a measure of distance, was calculated using a variation of a computer graphics method for calculating the shape distance between product designs (Kazhdan et al. 2003). The algorithm represents each 3D design based on spherical harmonics, in order to obtain rotation and to scale invariant characterizations that can be used to calculate distances that represent changes in shape rather than changes in perspective. One way to conceptually understand the technique is to imagine hollow 3D objects, and consider filling these objects with some number of tennis balls, ping pong balls, and ball bearings. Objects that are similar will need a similar proportion of balls of different sizes to fill them up. For Hypothesis 4, distance matrices were created between all pairs of parent designs and designs that reused them, making it possible to determine how dissimilar a new design was from its predecessor. The distance from the most *similar reuse* of the design (the closest child) and the distance from the most *dissimilar reuse* (the farthest child) were used as dependent variables for Hypothesis 4.

### Control Variables

A time-related control variable related to the designer, *designer tenure*, was included to control for the possibility that designs by long-standing members would be reused more (Faraj et al. 2015). *Designer tenure* was operationalized as the number of days between the first design contribution of the designer and the day that the design being analyzed was shared. Our second control variable, *design availability*, was also time-related and controlled for the possibility that designs that were available for a longer period would be reused more. *Design availability* was measured by the number of days that a design was available in the community for others to reuse. The time-related variables were log transformed due to skewed distributions.

Our last control variable was categorical and was included to control for the possibility that the format of representations of the designs could affect their reuse by others. We included the most common file types in Thingiverse, STL and OpenSCAD. Each of the designs was classified as *STL*, when it included only an STL version of the design, *OpenSCAD*, when designers had uploaded only an OpenSCAD script of the design, or *Both*. This usually happened when designers posted both the OpenSCAD and a rendering of the script in STL format. There was a fourth category, *Other*, that included different file types less common within the community. Hypothesis 4 also included *reuse* as a control variable. All control and independent variables were normalized by scaling between zero and one.

### Independent Variables

Two binary variables are used to indicate a design's relationship to customization. *Metamodel* indicates that the design is available as a functioning metamodel. All metamodels have an underlying OpenSCAD representation, but not all OpenSCAD files have interfaces that facilitate metamodel manipulation. *Generated* indicates that a design was created as the output of a metamodel. All of these files are in STL format, but not all STL format files are the result of customization. We measured the designers' *community experience* by counting the number of prior design contributions made by the designer, in line with other studies that measure experience in the form of contributions to the community (Crowston et al. 2012; Hann et al. 2013; Ransbotham and Kane 2011). The interaction of *metamodel* with *community experience* was used in Hypothesis 3. The means and correlations of all variables appear in Table A1 of the Appendix for the zero inflated negative binomial model (Hypotheses 1–3) and in Table A2 for the multiple regression model (Hypothesis 4).

### Modeling Approach

Our dependent variable is a count of reuse instances that are overdispersed, similar to activity distributions in most online community platforms (Shirky 2008), suggesting that a negative binomial model is appropriate. Very few of the designs contributed are reused even once. To counter the effects of excessive zeros in our model, we used a series of zero-inflated negative binomial regressions (Greene 1994).

Zero-inflated negative binomial regression assumes that there are two separate latent groups: (1) designs that have a nonzero likelihood of being reused and (2) designs that are not reused at all. The counts are generated by two separate processes to reflect the low probability that a design would be reused. The zero-inflated negative binomial regression model allows each observation to have a positive probability of being part of either group. The first process generates positive counts whereas the second process generates only zero counts. Therefore, two separate models are used to account for the two distinct latent processes. First, a binary logit model, also called an inflation model, is used to regress the zeroes (designs that will not be reused) and then a negative binomial regression model is used to regress the number of times a design is reused.

We also used a series of multiple regression models to test both parts of Hypothesis 4. The dataset included 813 designs that had been reused at least once. We used the same control variables that we used to test Hypotheses 1–3. In addition, we included *reuse* as a control variable, to account for the likelihood that a design would lead to more similar/dissimilar designs simply due to the number of times it was reused.

## Results

We performed a series of these zero-inflated negative binomial regressions for Hypotheses 1–3. Variables were added in a step-wise fashion to the models. In addition, we performed a number of additional analyses for robustness checks as well as to test the appropriateness of the analysis procedure. Negative binomial models are preferred over Poisson models when there is evidence of overdispersion. In this dataset, true dispersion was greater than 1, suggesting overdispersion. Dispersion estimate was 4.28 (p-value < 0.001). Vuong tests for all models suggested that a zero-inflated negative binomial regression model was a better fit for the data than a non-nested standard negative binomial (lowest z-stat 2.19, p-value = 0.01). The results are shown in Table 1 with design reuse as the dependent variable. Model 1 included both control variables related to time and *community experience*. The coefficients for prior *community experience* and *design availability* were positive and significant.

In Model 2 we examined the effects of the design representation formats. Designs that were shared solely in *OpenSCAD* format or included both *OpenSCAD* and *STL* representations had a positive significant relationship with design reuse, in contrast to the STL format. These results suggest that *OpenSCAD* is a more attractive format than *STL* when it comes to design reuse, and the next regression may explain why.

Model 3 included the *generated* flag in the zero-inflation model to test the generated design hypothesis (H2) that generated designs are less likely to be reused than designs that are not generated. The large positive value in the zero-inflation model suggests that *generated* designs were significantly associated with the likelihood of no further reuse. By contrast, the *metamodel* variable in the count model is strongly positive, suggesting a positive relationship with design *reuse*, providing support for the metamodel hypothesis (H1). It is also important to note that the *OpenSCAD* format variables are reduced in power compared to Model 2, suggesting that the strongest relationship of reuse is with the *metamodel* attribute. *STL* format was positive and marginally significant in Models 3 and 4, after factoring out the customized designs that are unlikely to have any reuse.

Finally, Model 4 tested the designer experience hypothesis (H3), the interaction between the *metamodel* and *community experience*. The interaction is positive and significant, suggesting that *community experience* positively moderates the positive effect of the *metamodel*.

We also controlled for possible correlations that could affect the significance of our results. Designs might be reused more because of the skills acquired by the designers outside the community (thus not captured from our *community experience* variable). In addition, certain categories of designs might tend to be reused more than others.

Using creator and category data, we calculated clustered standard errors to control for potential intraclass correlations, a technique that is used in econometrics (Cameron and Miller 2015). Our dataset contained 8,079 unique designers, 10 main design categories, and 79 design subcategories. We ran our zero-inflated negative binomial regressions and report results based on designer clustered standard errors (Table 1). The minor differences between (1) normally reported standard errors, (2) designer clustered standard errors, (3) category clustered standard errors, and (4) subcategory clustered standard errors do not significantly alter our results in general and our hypotheses in particular.

The most noticeable differences between the designer clustered standard error and a normally reported standard error were related to the *community experience* variable, as both were related to designer skills. In addition, we used the Walk-

**Table 1. Zero Inflated Negative Binomial Regression for Reuse**

| | | Model 1 | Model 2 | Model 3 | Model 4 |
|---|---|---|---|---|---|
| **Count Model** | | | | | |
| | Constant | -3.82*** | -7.18*** | -6.60*** | -6.58*** |
| | Community Experience | 5.58*** | 2.77** | 3.01*** | 1.92 |
| Control | Designer Tenure (ln) | 0.87 | 0.86*** | 0.37$^{\dagger}$ | 0.39$^{\dagger}$ |
| Control | Design Availability (ln) | 3.66*** | 3.76*** | 3.23*** | 3.23*** |
| Control | OpenSCAD | | 5.20*** | 1.21*** | 1.24*** |
| Control | STL | | -0.09 | 0.44$^{\dagger}$ | 0.45$^{\dagger}$ |
| Control | Both | | 4.76*** | 1.41*** | 1.43*** |
| H1 | Metamodel | | | 4.50*** | 4.34*** |
| H2 | Generated | | | -0.12 | -0.10 |
| H3 | Metamodel * Experience | | | | 3.25** |
| | *Log(θ)* | -3.29*** | -2.38*** | -1.52*** | -1.52*** |
| **Zero-Inflation Model** | | | | | |
| | Constant | 4.67*** | 1.61 | -8.81$^{\dagger}$ | -10.16* |
| | Community Experience | 2.00 . | 3.56 | 16.27*** | 17.43** |
| | Designer Tenure (ln) | -2.32*** | -1.12 | -3.29** | -3.14** |
| | Design Availability (ln) | -3.95*** | -4.20*** | -0.36 | -0.87 |
| H2 | Generated | | | 13.36*** | 15.10** |
| | DF | 9 | 12 | 15 | 16 |
| | $\theta$ | 0.04 | 0.09 | 0.22 | 0.22 |
| | Log-likelihood | -5,816 | -4,834 | -4,272 | -4,266 |
| | Wald $\chi^2$ | 48*** | 1,265*** | 1,630*** | 1,753*** |
| | **Adjusted pseudo $R^2$** | **0.06** | **0.21** | **0.31** | **0.31** |

N = 24,173; ***p < 0.001; **p < 0.01; *p < 0.05; $^{\dagger}$p < 0.10

trap community detection algorithm on the inheritance graph of designs, a null-model based procedure for clustering networks to produce the structures (Pons and Latapy 2006). Clustered standard errors based on the design families identified show only minor differences, and thus indicate that our findings are robust to conflation due to relatedness of designs.

For Hypothesis 4, we performed a series of multiple regressions with designer clustered standard errors; the results are shown in Table 2. Variables were added in a step-wise fashion similar to the way they were added in the zero-inflated negative binomial regression model. Models 5 and 9 included *reuse* as a control variable besides the ones used in Model 1.

The effect of design representation formats (*OpenSCAD*, *STL,* or *Both*) was strong and significant for both minimizing the distance between the parent design and its *most similar reuse* (Models 6 and 7), and for maximizing the distance between the parent design and its *most dissimilar reuse* (Models 10 and 11). Models 7 and 11 included the *metamodel* variable to test the Hypothesis 4 as well as the *generated* variable. *Metamodel* had a negative significant effect on the distance between the parent design and its *most similar reuse*, and positive significant effect on the distance between the parent design and its *most dissimilar reuse*. Finally, Models 8 and 12 included the interaction between the *metamodel* and *community experience*. Table 3 summarizes the results.

As a robustness check for the impact of the *metamodel* on design *reuse*, and in order to make sure we were not seeing the effects of pure substitution, we looked at 120 designs that were uploaded on Thingiverse before the introduction of the customizer and were later updated to a customizer version. These designs were reused more after the introduction of the customizer ($\mu_{differences}$ = 1.82, paired t-test p-value = 0.003), even though they were available as non-customizers for signi-

**Table 2. Multiple Regressions for Shape Distance**

| | | Model 5 | Model 6 | Model 7 | Model 8 | Model 9 | Model 10 | Model 11 | Model 12 |
|---|---|---|---|---|---|---|---|---|---|
| | | Distance to most similar reuse | | | | Distance to most dissimilar reuse | | | |
| | Constant | 0.57*** | 0.84*** | 0.84*** | 0.85*** | 0.53*** | 0.69*** | 0.68*** | 0.67*** |
| | Community Experience | 0.17** | 0.06 | 0.06 | 0.04 | -0.13† | -0.04 | -0.04 | -0.00 |
| Contr | Designer Tenure (ln) | -0.05* | 0.00* | 0.00 | -0.00 | 0.08** | 0.04† | 0.03 | 0.04† |
| | Design Availability (ln) | -0.12** | -0.09** | -0.09* | -0.09* | 0.10* | 0.07† | 0.08† | 0.07† |
| | Reuse | -1.05** | -0.82*** | -0.78** | -0.85*** | 1.13** | 0.93** | 0.86** | 0.99** |
| | OpenSCAD | | -0.39*** | -0.29*** | -0.30*** | | -0.05* | -0.22*** | -0.22*** |
| | STL | | -0.23*** | -0.23*** | -0.24*** | | -0.18*** | -0.18*** | -0.18*** |
| | Both | | -0.37*** | -0.29*** | -0.29*** | | -0.06*** | -0.21*** | -0.21*** |
| | Metamodel | | | -0.10*** | -0.10*** | | | 0.19*** | 0.19*** |
| | Generated | | | 0.03 | 0.03 | | | 0.00 | 0.00 |
| | Metamodel*Experience | | | | 0.10 | | | | -0.19 |
| | DF | 808 | 805 | 803 | 802 | 808 | 805 | 803 | 802 |
| | F-stat | 25.72*** | 36.81*** | 31.44*** | 28.39*** | 27.57*** | 29.81*** | 31.12*** | 28.37*** |
| | **Adjusted $R^2$** | **0.11** | **0.24** | **0.26** | **0.26** | **0.12** | **0.21** | **0.26** | **0.26** |

N = 24,173; ***p < 0.001; **p < 0.01; *p < 0.05; †p < 0.10

**Table 3. Summary of Findings**

| | Hypothesis | Finding |
|---|---|---|
| H1 | Metamodels are more likely to be reused than models. | Supported |
| H2 | Designs that are generated from metamodels are less likely to be reused than other designs. | Supported |
| H3 | Metamodels will exhibit amplified reuse when created by members with higher levels of community experience. | Supported |
| H4 | Metamodels are more likely than models to lead to designs similar to themselves, and therefore are less likely to lead to dissimilar designs. | Partially Supported |

ficantly fewer days ($\mu_{differences} = 157.22$, paired t-test p-value < 0.001). A Wilcoxon rank sum test also confirmed these results (p-value < 0.001). Thus our results are robust against that alternative explanation. ANCOVA models were used as robustness checks to determine the statistical significance of the effect of the *metamodel* on design similarity, while controlling for the factors mentioned above. There was a significant effect of *metamodel* on most *similar reuse* $F(1, 802) = 19.32$, $p < 0.001$, as well as on most *dissimilar reuse* $F(1, 802) = 57.26$, $p < 0.001$.

## Discussion and Concluding Thoughts ■

Theories of knowledge reuse point out that knowledge repositories attract both expert practitioners and expertise-seeking novices (Markus 2001). Moreover, they make a distinction between reuse for replication and reuse for innovation (Majchrzak et al. 2004). These theories were tested on reuse data extracted from an open online design community. This community introduced a metamodel they called a customizer, a technology that has been studied in consumer-oriented innovation contexts. We hypothesized that reuse from a metamodel constituted a different form of reuse, distinct from reuse for replication and reuse for innovation. Metamodels were reused often, but the generated models were not reused. Metamodels led to the creation of more similar designs when compared to models (H4). But metamodels also led to more dissimilar designs. One interpretation of this result is that designers used the metamodels to perform usually local but occasionally more distant searches. Indeed, Figures A1 and A2 of the Appendix imply this is the case. What more can we determine about the mechanisms at work?

The analysis suggests that much of reuse variance can be explained based on choices that designers make about formats and tools. In particular, metamodels can be reused in two ways: they can be run in order to generate customized models, or they can be extended as new metamodels. The interaction effect in Model 4 suggests that metamodels created by experienced designers are particularly likely to be reused. Perhaps metamodels are usually the result of a process of modification by experienced designers, and generated models are typically the results of experiments by novice designers.

Indeed, those generating customized models had lower tenure in the community ($\mu_1 = 53.68$ versus $\mu_2 = 139.86$, p-value < 0.001). They also contributed fewer designs before the introduction of the customizer ($\mu_1 = 5.91$ versus $\mu_2 = 11.74$, p-value < 0.001). Designers who contributed only generated models were 41.4% of the total user population, as opposed to 48.6% contributing only models or metamodels, with 9.9% contributing both. Designers that contributed only generated models created only 0.6% of all designs that were reused (p-value < 0.001). It looks like experienced designers largely created the metamodels, and novices ran the metamodels. But a large portion of designers that created metamodels also ran them in order to generate specific variants of the design (32%). We also saw some examples of designers who generated models early in their tenure and ended up learning Blender or OpenSCAD, and building metamodels themselves.

The introduction of the metamodel had a profound effect on the community. Thingiverse had 28,774 designs created over a four year period before the platform introduced the metamodel tool they called the customizer. After that, they crossed the 100,000 object milestone in just six more months (Makerbot Blog 2013). In response to a blog post making this observation, one designer wrote

> *Are there stats on how many of the 100,000 are just slightly different useless custom variations of something and how many are actual unique things?*

Another wrote

> *The customizer is great, but you really need to adapt the search so that I can remove all the customized things that people put up there. At this point it's almost useless.*

Thingiverse increased overall reuse, but risked alienating experienced community members. Even if they were annoyed, experienced designers did create metamodels, as shown by our test of the designer experience hypothesis. Previous literature on knowledge repositories portrayed such sharing of explicit knowledge as being hard to incentivize (Markus 2001; Orlikowski 1992). Perhaps the high reuse that accompanies metamodels provides a quick injection of positive feedback to the designer, which encourages the creation of more metamodels.

Our findings have a practical implication for platform managers: because there are different processes of reuse, it could be helpful to provide technological features that support different processes of search for different reuse objectives. Search for customization might help novices find a metamodel from which they can generate a model. Search for innovation might screen out all generated designs, and even suggest categories of objects where new metamodels might be useful. How might this be done? While this analysis has focused on variations in shape, it will be important to know if the search for novel form in these communities also leads to novel function (Frenken 2006; Saviotti 1996). Examining the interaction between form and function differences of designs could provide further insights about how artifacts are conceived, developed, used, and reused. Then it might be possible to highlight for designers sets of designs where form and function seem to be changing: this may be the frontier of innovation. Metamodels may first serve as exploratory tools for pressing out the frontier, and later serve as tools for consolidating what the community learned.

Metamodels exist in other domains too. For example, ERP systems vendors introduced metamodels to reduce large-scale tailoring of their systems (Sarker et al. 2012). Our findings suggest that users in other domains will be pulled toward using the metamodels, not generated models. Architecture is also making use of digital technologies to customize buildings (Berente et al. 2010; Boland et al. 2007). In particular, architectural systems that model information about buildings utilize metamodels. There may be a tradeoff involved in attracting and supporting less experienced workers without alienating highly experienced workers. If standardization and openness in professional design increases, we might see expert architects creating metamodels for use by themselves to explore design space and for use by expertise-seeking novices.

Our findings also have implications for software reuse, a specific form of knowledge reuse studied in information systems (Allen and Parsons 2010; Karimi 1990; Kim and Stohr 1998; Purao et al. 2003; Sherif et al. 2006). Research conducted inside a large technology company noted that the best division of labor occurs when experts create components that are reused by novices (Lim 1994). We found that experts gravitated toward the metamodels, and that the metamodels were more reused when built by experts. Unlike that company's situation, in which engineers were part of a hierarchy, on Thingiverse this division of labor happens naturally through a process of self-selection. Likewise, obstacles to software reuse found in companies demand continuous managerial intervention (Sherif et al. 2006). In open design communities, these interventions are not possible, nor do they appear to be necessary; the issues in such communities are less about motivating reuse and more about making the reuse productive (Benkler et al. 2015; Hill and Monroy-Hernández 2012, 2013).

Information systems scholars have noted that reuse is not always a good thing; because of anchoring, errors can be introduced (Allen and Parsons 2010). In an open design community, anchoring is also likely to occur, but it is our conjecture that its deleterious effects are also likely to be quickly fixed, either by the designer who finds the object cannot be 3D printed, or by others who offer their own alternatives.

Information systems scholars have also explored software metamodels in the form of computer-aided software engineering (Banker and Kauffman 1991; Jarke et al. 2009; Orlikowski 1993). There are many layers possible: languages can be generated by other languages (models, metamodels, metametamodels) to as many levels as are desired (Jarke et al. 2009). But universal modeling tools have had slow adaption. By contrast, some tools focused on particular domains have been successful (Kelly and Tolvanen 2008). In the open design community we studied, each metamodel is like a domain-specific language, defining a family of designs. Would this work in an open software design environment? Our findings suggest the possibility of an open software community in which the domain-specific models, based on a common platform, can themselves be easily extended as well as easily run.

Is there another possible higher metalevel for open design and software environments? It might be productive to permute defaults in the metamodels to take advantage of any anchoring tendency. That is, by changing defaults to cover relatively unexplored parts of the environment, or by changing the visibility of reusable components, the collective might be nudged to explore more of the search space.

These conjectures suggest a conceptualization of reuse that considers a spectrum of reuse practices, including replication, customization, and radical innovation. In open communities, members engage in local search through customizing designs, and exploration through the modifications and recombinations that yield not just individual designs but also families of designs. Evolutionary models of diffusion (Arakji and Lang 2010), as well as theories of market microstructure, provide possible beginning points for such a conceptualization of reuse (Holzmann et al. 2014).

Open design communities may turn these observations into actions. Tools can be designed to provide ways to search for designs that might be most extensible, designs that are not as far along in their trajectories of development, and have not stabilized on a family of designs. They might also suggest the recombination of unusual pairings of stable families of designs, leading to novel and practical designs, the ultimate goal of reuse for innovation.

While customization is a subject of current interest (Fogliatto et al. 2012; Franke 2016; Piller and Salvador 2016), it is easy to forget that a few generations ago it was the status quo. Customization virtually disappeared as a result of the standardization movement that launched a century ago (Lampel and Mintzberg 1996; Noble 1979; Yates and Murphy 2015). Many items, such as shoes, that previously were designed to fit individuals were instead issued in standard sizes in order to take advantage of economies of scale (Alford 1929; Lampel and Mintzberg 1996). In contrast, the newer technology discussed here allows for each object to be of a different size, with little additional cost (Conner et al. 2014; Huang et al. 2013).

The confluence of the digital and physical in 3D printing technologies brings us back to customized manufacturing. But, because of the nature of the digital, this is a more affordable and more rapid form of customization than in the past. For information systems theory and practice, the confluence of the digital and physical is a largely unexplored territory worth exploring, as it has the potential to fundamentally change our environment.

## Acknowledgments

The authors thank the editors and reviewers for their direction. We thank Steven Englehardt and Joseph Risi for the collection and analysis of data, as well as the members of the Center for Decision Technologies at Stevens Institute of Technology for their feedback. We also thank Sinan Aral for a helpful discussion on network autocorrelation. This material is based upon work supported by the National Science Foundation under grants IIS-1211084, IIS-1422066, and CCF-1442840.

## About the Authors

**Harris Kyriakou** is an assistant professor in Information Systems at IESE Business School. His research interests include collective innovation, computer-supported cooperative work, and crowdsourcing. He focuses on the evolution of digital artifacts from a social and information network perspective. He also studies the parameters of innovation processes.

**Jeffrey V. Nickerson** is a professor in the School of Business at Stevens Institute of Technology. His research interests include crowd work, collective intelligence, and design. Prior to joining Stevens, he was a partner at PricewaterhouseCoopers, where he consulted on issues related to software design and development. He holds a Ph.D. in Computer Science from New York University.

**Gaurav Sabnis** is an assistant professor in the School of Business at Stevens Institute of Technology. His research interests include online user-generated content, social media and sales. He has published in *Journal of Marketing, Information Systems Research*, and *MIS Quarterly*, among other journals. He holds a Ph.D. in Marketing from Penn State University.

# Appendix

**Table A1. Means and Correlations for Hypotheses 1 through 3**

| | Mean | s.d. | 1 | 2 | 3 | 4 | 5 | 6 | 7 | 8 | GVIF |
|---|---|---|---|---|---|---|---|---|---|---|---|
| 1. Community Experience | 0.02 | 0.08 | | | | | | | | | 1.16 |
| 2. Designer Tenure (ln) | 0.33 | 0.31 | 0.36*** | | | | | | | | 1.21 |
| 3. Design Availability (ln) | 0.78 | 0.20 | 0.03*** | 0.02* | | | | | | | 1.00 |
| 4. OpenSCAD | 0.01 | 0.12 | 0.01* | 0.07*** | 0.04*** | | | | | | |
| 5. STL | 0.87 | 0.33 | -0.02* | -0.14*** | -0.06*** | -0.31*** | | | | | 2.17 |
| 6. Both | 0.07 | 0.25 | 0.04*** | 0.18*** | 0.05*** | -0.03*** | -0.70*** | | | | |
| 7. Metamodel | 0.03 | 0.17 | 0.03*** | 0.12*** | 0.07*** | 0.45*** | -0.44*** | 0.40*** | | | 2.07 |
| 8. Generated | 0.45 | 0.50 | -0.12*** | -0.28*** | -0.13*** | -0.11*** | 0.31*** | -0.22*** | -0.15*** | | 1.02 |
| 9. Reuse | 0.45 | 11.74 | 0.08*** | 0.04*** | 0.03*** | 0.10*** | -0.10*** | 0.09*** | 0.21*** | -0.03*** | — |

N = 24,173; ***p < 0.001; **p < 0.01; *p < 0.05

**Table A2. Means and Correlations for Hypothesis 4**

| | Mean | s.d. | 1 | 2 | 3 | 4 | 5 | 6 | 7 | 8 | 9 | 10 | GVIF |
|---|---|---|---|---|---|---|---|---|---|---|---|---|---|
| 1. Community Experience | 0.04 | 0.10 | | | | | | | | | | | 1.26 |
| 2. Designer Tenure (ln) | 0.51 | 0.33 | 0.41*** | | | | | | | | | | 1.14 |
| 3. Design Availability (ln) | 0.87 | 0.13 | 0.04 | -0.01 | | | | | | | | | 1.01 |
| 4. Reuse | 13.24 | 62.67 | 0.33*** | 0.12*** | 0.06 | | | | | | | | 1.31 |
| 5. OpenSCAD | 0.17 | 0.38 | 0.01 | 0.02 | 0.01 | 0.08* | | | | | | | |
| 6. STL | 0.40 | 0.49 | 0.03 | -0.17*** | -0.07* | -0.16*** | -0.37*** | | | | | | 1.26 |
| 7. Both | 0.42 | 0.49 | -0.04 | 0.15*** | 0.06 | 0.10** | -0.39*** | -0.71*** | | | | | |
| 8. Metamodel | 0.52 | 0.50 | 0 | 0.16*** | 0.05 | 0.18*** | 0.41*** | -0.85*** | 0.53*** | | | | 2.02 |
| 9. Generated | 0.01 | 0.11 | -0.02 | -0.01 | -0.02 | -0.02 | -0.05 | 0.07 | -0.03 | -0.12*** | | | 1.01 |
| 10. Similar Reuse | 0.44 | 0.19 | -0.04 | -0.07* | -0.11** | -0.31*** | -0.19*** | 0.41*** | -0.27*** | -0.44*** | 0.06 | | — |
| 11. Dissimilar Reuse | 0.66 | 0.21 | 0.08* | 0.13*** | 0.09* | 0.32*** | 0.16*** | -0.36*** | 0.23*** | 0.44*** | -0.05 | -0.06 | — |

N = 813; ***p < 0.001; **p < 0.01; *p< 0 .05

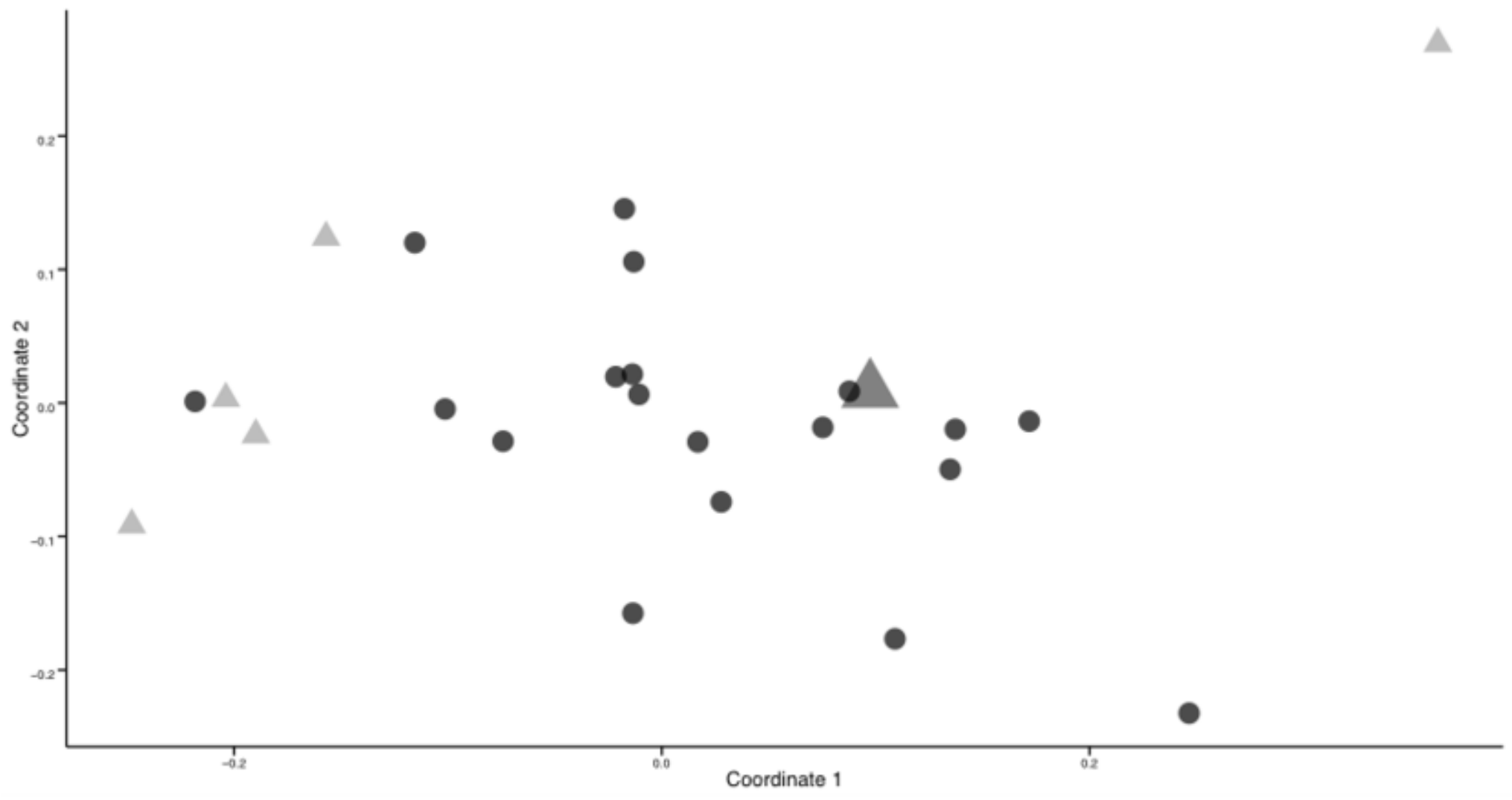

**Figure A1. Multidimensional Scaling Representation of the Children of a Metamodel (The metamodel is shown as a large triangle. Circles indicate generated models and small triangles indicate edited metamodels.)**

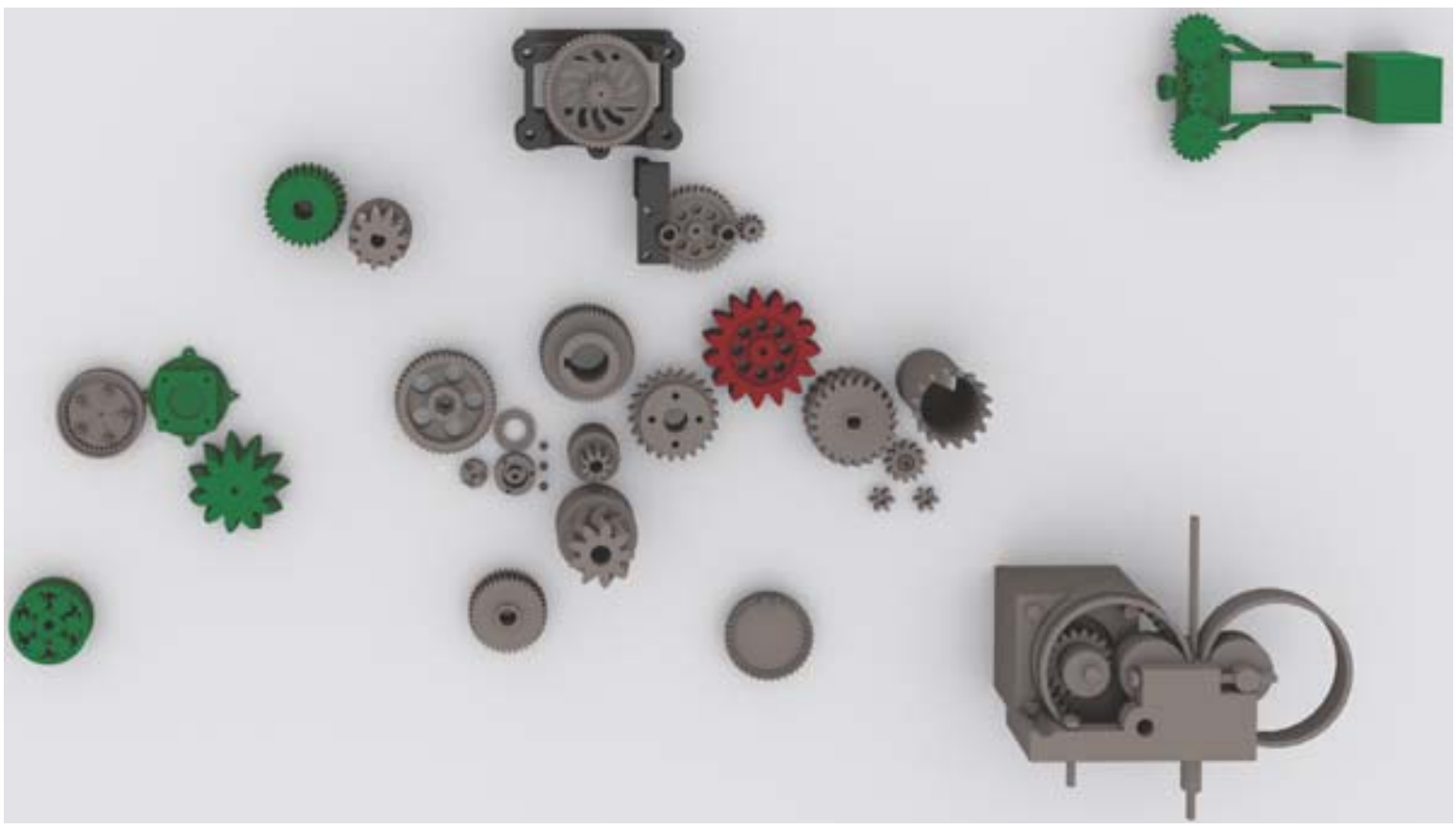

**Figure A2. Example Parts from the Designs Placed According to the Multidimensional Scaling Placements of Figure A1.**